\documentclass[12pt]{article}
\begin{document}
\pagestyle{empty}

\begin{center}
{\Large \bf
Infrared renormalons and 
the relations between the Gross--Llewellyn Smith and     
the Bjorken polarized and  unpolarized sum rules}

\vspace{0.1cm}

{\bf A.L. Kataev}\footnote{E-mail:kataev@ms2.inr.ac.ru }\\
\vspace{0.1cm}

Institute for Nuclear Research of the Russian  Academy of Sciences
,\\ 117312  Moscow, Russsia

\end{center}
\begin{center}

{\bf  Abstract}
\end{center}
It is demonstrated that the infrared 
renormalon calculus indicates that the QCD  theoretical 
expressions for the Gross--Llewellyn Smith sum rule and for the Bjorken 
polarized and unpolarized ones 
contain an identical negative twist-4 $1/Q^2$ correction.
This observation is supported by the consideration 
of the results of calculations of the corresponding 
twist-4 matrix elements. Together with the 
indication of the  similarity of perturbative QCD 
contributions to these three sum rules, this 
observation  leads to simple new theoretical  relations 
between the Gross-Llewellyn Smith 
and Bjorken polarized and unpolarized sum rules in the energy region
$Q^2\geq$ 1 ${\rm GeV}^2$. The validity of this relation is checked 
using concrete experimental data for the Gross--Llewellyn Smith 
and Bjorken polarized sum rules.

\noindent 
\vspace*{0.1cm}
\noindent
\vspace*{2cm}
\vfill\eject

\setcounter{page}{1}
\pagestyle{plain}

It is known that,  in the traditionally used $\overline{\rm MS}$ scheme, 
the Borel image of the physical quantities in QCD  contain 
infrared renormalons
(IRR), namely the singularities on the positive axis of integration of this 
image in the complex plane of the  Borel variable $\delta$ (for reviews 
see Refs. \cite{Beneke:1998ui}--\cite{Altarelli:1995kz}.
This related  Borel integral can be  defined as 
\begin{equation}
D(a_s)=\int_0^{\infty} {\rm exp}(-\delta/{\beta_0 a_s})B[D](\delta)d\delta~,
\end{equation}
where $a_s=\alpha_s/(4\pi)$; $\alpha_s$ is the QCD coupling constant 
in the $\overline{\rm MS}$ scheme;  $\beta_0=(11/3)C_A-(4/3)T_fN_f$ 
is the first coefficient of the 
QCD $\beta$-function with  $C_A=3$, $T_f=1/2$;  
and $B[D](\delta)$ is the Borel image 
of the physical quantity $D(a_s)$ under consideration. 

From  our point of view, the most important theoretical works, 
which pushed ahead   
the study of the applicability  of the IRR calculus  
to the analysis of non-perturbative contributions to the  characteristics 
of  different processes 
are those 
of Ref. \cite{Zakharov:1992bx} and  \cite{Mueller:1993pa}.
In particular, it was shown  in Ref. \cite{Mueller:1993pa} that 
since there exist a 
$1/Q^2$ 
non-perturbative 
correction 
of twist-4 
in the characteristics of deep-inelastic scattering, 
the related Borel images should have the IRR pole at $\delta=1$; this  
does not manifest itself in the expression for the Borel image 
of  the Adler $D$-function of  the $e^+e^-$ annihilation process 
\cite{Zakharov:1992bx}.
This crucial  remark   later generalized 
to the discussion of 
the Bjorken polarized sum rule in Ref. \cite{Ji:1994md}. 
It should  also be  mentioned  in passing  that ultraviolet renormalons (UVR), 
associated with 
sign-alternating  asymptotic perturbative contributions to the
QCD perturbative series, 
manifest themselves in the Borel images as the poles 
at $\delta$ = - $k$, where $k$ are  integer numbers.

The next problem, which arises in  the process of applying  
the renormalon calculus to the analysis of the structure of both 
asymptotic perturbative 
contributions and non-perturbative corrections to physical quantities
is the calculation of the corresponding  Borel images $B[D](\delta)$.
These calculations are usually performed  using   large-$N_f$ expansion
(where $N_f$ is the number of quarks flavours). What is really 
calculated is 
the so-called one-renormalon-chain 
approximation to the  Born expression  
for  the  physical quantity under consideration.
Note that the  renormalon chain is associated 
with the gluon propagator, 
dressed by a large number 
by  quark 
bubbles insertions labelled 
by $N_f$.   
The contributions of this chain into the  theoretical   expression for  
physical quantities  are gauge-invariant, but they  do not reflect
the whole picture of renormalon effects in QCD. The latter begin  
to manifest themselves after application of the naive non-abelianization 
ansatz \cite{Broadhurst:1994se} only, namely after the 
replacement   $N_f\rightarrow -(3/2)\beta_0$ = $N_f-(33/2)$ in the leading 
terms of the large-$N_f$ expansion.
 This procedure  transforms a    
large-$N_f$ expansion  into  a  large-$\beta_0$ expansion,
also considered  in some  recent  works  \cite{Aquila:2005hq}, 
where it was associated with a  BLM-type expansion \cite{Brodsky:1982gc}.

In this Letter,  definite new  consequences of the 
relations 
between  the Borel images,  calculated 
in Ref. \cite{Broadhurst:1993ru}  and  \cite{Broadhurst:2002bi}   
for the  Gross--Llewellyn Smith (GLS) 
sum rule  of $\nu N$ deep-inelastic scattering (DIS) \cite{Gross:1969jf}, 
the Bjorken polarized (Bjp) sum rule \cite{Bjorken:1966jh} of 
polarized  charged-lepton--nucleon  DIS  
and the Bjorken unpolarized (Bjunp) sum rule \cite{Bjorken:1967px}
of $\nu N$ DIS are discussed. 
In particular, it is argued that the values 
of the matrix elements of the twist-4 $1/Q^2$ corrections 
to the Bjp, Bjunp  and GLS sum rule should have the same value. 
Together with the similarity in  the behaviour of 
the  perturbative corrections 
of  all three  sum rules, discussed  in Ref. \cite{Broadhurst:2002bi},
this new observation allows  us to write theoretical 
expressions to relate relate 
these  a priori different 
physical quantities. 

To be more precise, 
consider first  the definitions of the sum rules we are interested in, 
taking into account twist-4 operators, evaluated 
in Ref. \cite{Shuryak:1981kj} in the case 
of GLS and Bjunp sum rules and in Ref.\cite{Shuryak:1981pi}
for the  Bjp sum rule: 
\begin{eqnarray}
{\rm GLS}(Q^2)&=&\frac{1}{2}\int_0^1dx\bigg[F_3^{\nu n}(x,Q^2)+
F_3^{\nu p}(x,Q^2)\bigg]=3C_{\rm GLS}(Q^2)-\frac{\langle\langle 
O_1 \rangle\rangle}{Q^2}~~~, \\ 
{\rm Bjp}(Q^2)&=&\int_0^1dx\bigg[g_1^{lp}(x,Q^2)-g_1^{ln}(x,Q^2)]=
\frac{g_A}{6}C_{\rm Bjp}(Q^2)-\frac{\langle\langle O_2 \rangle\rangle}
{Q^2}~~~, \\
{\rm Bjunp}(Q^2)&=&\int_0^1dx\bigg[F_1^{\nu p}(x,Q^2)-F_1^{\nu n}(x,Q^2)\bigg]=
C_{\rm Bjunp}(Q^2)-\frac{\langle\langle O_3 \rangle\rangle}{Q^2}~~~.
\end{eqnarray}
where 
\begin{eqnarray}
C_{\rm GLS}&=&1-4a_s - O(a_s^2)~~~,\\
C_{\rm Bjp}&=&1- 4a_s - O(a_s^2)~~~,\\
C_{\rm Bjunp}&=&1-\frac{8}{3}a_s- O(a_s^2)~~~.
\end{eqnarray}
The explicit 
expressions of the numerators of twist-4 contributions 
are defined as in the  review  \cite{Hinchliffe:1996hc}, namely
\begin{eqnarray}
\langle\langle O_1 \rangle\rangle &=&\frac{8}{27}
\langle\langle O^{\rm s} \rangle\rangle~~~, \\
\langle\langle O_2 \rangle\rangle &=& \langle\langle O^{\rm NS}_{p-n} 
\rangle\rangle~~~, \\
\langle\langle O_3 \rangle\rangle &=&\frac{8}{9}\langle\langle O^{\rm NS}
\rangle\rangle~~~,
\end{eqnarray}
where matrix elements on the r.h.s. of Eqs.(8)--(10)
are known explicitly. 
Indeed, the matrix elements $\langle\langle O^{\rm s} \rangle\rangle$ and 
$\langle\langle O^{NS}\rangle\rangle$
were calculated in Ref. \cite{Shuryak:1981kj}, while the 
matrix element $\langle\langle O^{NS}_{p-n}\rangle\rangle$ is  calculated 
in Ref. \cite{Shuryak:1981pi}.

Let us return to renormalon calculus.
It is known from Ref. \cite{Broadhurst:1993ru} that the 
 Borel images for the GLS and Bjp sum rules coincide and have the following 
form:
\begin{equation}
B[C_{\rm GLS}](\delta)=B[C_{\rm Bjp}](\delta)= 
-\frac{(3+\delta) {\rm exp}(5\delta/3)}{(1-\delta^2)(1-\delta^2/4)}~~.
\label{GLS}
\end{equation}
It was shown  in Ref. \cite{Broadhurst:2002bi}, 
that the Borel image $B[C_{\rm Bjunp}](\delta)$ 
of the Bjunp sum rule 
is closely related to Eq. (11), namely 
\begin{equation}
B[C_{\rm GLS}](\delta)=\bigg(\frac{3+\delta}{2(1+\delta)}
\bigg)B[C_{\rm Bjunp}(\delta)]~~~,
\label{relation}
\end{equation}
where 
\begin{equation}
B[C_{\rm Bjunp}](\delta)=-
\frac{2{\rm exp}(5\delta/3)}{(1-\delta)(1-\delta^2/4)}~~~.
\label{Bjunp}
\end{equation}

The consideration of  Eqs. (\ref{GLS})--(\ref{Bjunp})
allow the following conclusions to be made  \cite{Broadhurst:2002bi}:
in the $\overline{\rm MS}$-scheme the asymptote  of perturbative series 
for the GLS, Bjp and Bjunp sum rules is dominated by the 
first $\delta=1$ IRR. Indeed, in the case of GLS and Bjp sum rules,  
the first UVR  at $\delta=-1$, responsible for the 
sign-alternating  perturbative QCD contribution to the asymptotic 
behaviour of the perturbative QCD series (for a  more detailed discussion see 
Refs. \cite{Beneke:1998ui,Beneke:2000kc}, is suppressed by a factor 
$(1/2){\rm exp}(-10/3)=0.018$, with respect  to the dominant IRR 
at $\delta=1$, 
responsible for sign-constant $n!$ growth  of the 
perurbative coefficients for these two sum rules. Moreover, it is 
obvious from the results of Ref. \cite{Broadhurst:2002bi}  that 
in the case of the  Bjunp sum rule the first UVR, 
created by the pole at $\delta=-1$, 
is absent and that the residues of the first IRR in the Borel images 
for the GLS, Bjunp and Bjp sum rules are the same.
Therefore, it is possible to make the conclusion  that the
asymptotic  perturbative QCD 
contributions will have an  identical structure \cite{Broadhurst:2002bi}.
This fact is supported by the  
next-to-next-to-leading order studies 
of Refs. \cite{Gardi:1998rf}, performed  
 with the help of the method of effective 
charges \cite{Grunberg:1982fw}.  

Now I will  make the new conclusion, which follows 
from the results of Eqs. (\ref{GLS})--(\ref{Bjunp})
and is related to twist-4 $O(1/Q^2)$ corrections. Since the  IRR 
contribution of the first $\delta=1$ IRR pole enter into  
the Borel images of the GLS, Bjp and Bjunp sum rules 
with the same negative residue  (see (\ref{GLS})--(\ref{Bjunp})),
the  normalized to unity  $O(1/Q^2)$ power correction in 
the GLS, Bjp and Bjunp sum rules, which are  related to  the 
 $O(\Lambda^2/Q^2)$ ambiguities  
in the   Borel integrals  generated by the  $\delta=1$ IRR pole,    
should have the same sign and a  similar 
numerical value. Indeed, the  $\Lambda^2/Q^2$ IRR ambiguities  
may be coordinated with the definitions of the twist-4 contributions
(see e.g. the reviews \cite{Beneke:1998ui,Beneke:2000kc}) and if they 
are the same the twist-4 $1/Q^2$ corrections should be the same 
also.

Let us check this statement, using concrete results 
of calculations of the    numerical values of the matrix elements 
by means of three-point function QCD sum rules, namely 
$\langle\langle O^s\rangle\rangle=0.33~ 
\rm{GeV}^2$, $\langle\langle O^{\rm NS}\rangle\rangle = 0.15~\rm{GeV}^2$  
obtained in 
Ref. \cite{Braun:1986ty}. As to the error bars, we propose 
to use $50\%$ conservative uncertainty.
This choice is in agreement with 
the error bar of the following value for 
twist-4 matrix element for the Bjorken polarized sum rule,
namely  $\langle\langle O^{\rm NS}_{p-n}\rangle\rangle = 0.09 \pm 0.06~ 
{\rm GeV^2}$, obtained in Ref. \cite{Balitsky:1989jb}. Taking 
now $g_A=1.26$, we find the following 
numerical expressions for the sum rules of Eqs.(2)--(4):
\begin{eqnarray}
{\rm GLS}(Q^2)& =&3\bigg[1-4a_s-O(a_s^2) -
\frac{0.098~{\rm GeV^2}}{Q^2}\bigg]~~~, \\
{\rm Bjp}(Q^2)&=
&\frac{g_A}{6}\bigg[1-4a_s-O(a_s^2)-\frac{0.071~{\rm GeV^2}}{Q^2}
\bigg]~~, \\
{\rm Bjunp}(Q^2)&=&\bigg[1-\frac{8}{3}a_s -
O(a_s^2)-\frac{0.133~{\rm GeV^2}}{Q^2}\bigg]~~~.
\end{eqnarray}
One can see that, within  theoretical uncertainties, typical 
of the  application of three-point function QCD sum rules, 
the prediction of the IRR calculus 
is confirmed. So, indeed, the  $O(1/Q^2)$ corrections normalized to unity, 
have the same negative sign and   very  close   values. 

In view of the fact that IRR calculus also indicates that known and 
still unknown   perturbative 
QCD corrections to all three sum rules  have  comparable value 
as well \cite{Broadhurst:2002bi}, I  now  write  the following relation
between the three sum rules we are interested in, namely 
\begin{equation}
{\rm Bjp}(Q^2) \approx (g_A/18){\rm GLS}(Q^2)\approx 
(g_A/6){\rm Bjunp}(Q^2)~~~.
\label{basic}
\end{equation}
These  relations are  valid in the energy region where 
it is possible to separate the twist-4 contribution from the  twist-2 effects 
and $1/Q^4$ contributions are smaller than $1/Q^2$ effects. 
The above-mentioned features should hold  at    $Q^2\geq$ 1 ${\rm GeV^2}$.
Therefore, theoretical  comparisons  \cite{Shirkov:2001sm} 
of the expressions for the Bjp sum rule \cite{Milton:1998cq} and the  
GLS sum rule \cite{Milton:1998cq} within analytic approch \cite{Shirkov:1997wi}
should possess the same feature.

Now I will consider  whether the l.h.s. of the  basic equation (\ref{basic})
is respected by experiment. I will use 
the values for the GLS sum rule,  extracted in Ref. \cite{Kim:1998ki}, 
at the energy points  
$Q^2$=1.26 GeV$^2$, 2 GeV$^2$, 3.16 GeV$^2$, 5.01 GeV$^2$,
7.94 GeV$^2$, 12.59 GeV$^2$. The results presented in Ref. \cite{Kim:1998ki}  
for these six energy points are  ${\rm GLS}(Q^2) \approx$ 
2.39, 2.49, 2.55, 2.78, 
2.82 and 2.80, where for simplicity we neglected both statistical 
and systematical uncertainties. 
The application of the l.h.s. of Eq. (\ref{basic})
gives one the following experimentally motivated values 
for  the Bjp sum rule, namely $\rm{Bjp}(Q^2) \approx$ 0.167, 0.168, 
0.178, 0.195, 0.197, 0.196 for the same energy points, where again 
the contribution of statistical and systematical uncertainties are 
not taken into account.

It is interesting that the value of the Bjp sum rule 
extracted in Ref. \cite{Altarelli:1996nm} 
from the SLAC and SMC data 
is ${\rm Bjp}(Q^2=3~{\rm GeV^2)}=
0.177\pm 0.018$  and,  within existing error bars,  do not contradict 
the value ${\rm Bjp}(Q^2=3~{\rm GeV^2})=0.164\pm 0.011$ extracted 
in Ref. \cite{Ellis:1995jv} on the basis of measurements at CERN and SLAC 
before 1997. It is rather inspiring that these results 
agree with the GLS sum rule value at $Q^2=3.16 ~{\rm GeV}^2$. 

At relatively  high  energies 
the SMC 
collaboration gives ${\rm Bjp}(Q^2=10~{\rm GeV}^2)=0.195\pm 0.029$ 
\cite{Adeva:1997is} which is consistent with high-energy results 
for the GLS sum rule ${\rm GLS}(Q^2=12.59~{\rm GeV}^2) \approx 
0.196$ 
\cite{Kim:1998ki}. However, at low $Q^2$
the result ${\rm Bjp}(Q^2= 1.10~{\rm GeV^2}) \approx 0.136$, 
extracted from CEBAF data
in Ref. \cite{Deur:2004ti}, is not consistent with the estimate
${\rm Bjp}(Q^2 =1.26~{\rm GeV^2}) \approx 0.167$ extracted 
low-energy results ${\rm GLS}(Q^2=1.26~{\rm GeV^2})\approx 2.39$ 
\cite{Kim:1998ki} with the help of 
Eq. (\ref{basic}).
It may be interesting to clarify the origin of this disagreement, taking  
into account experimental uncertainties of the   two independent 
analyses of $\nu N$ DIS data and $lN$ polarized DIS data. 
As the next step one could  
check the consistency of other experimental results for the GLS sum rule 
and Bjp sum rule with the IRR motivated expression of 
 Eq. (\ref{basic}) for the  energy points 
in the region $1~{\rm GeV}^2\leq Q^2 \leq 5~{\rm GeV}^2$ using 
NuTeV data for $xF_3$ structure function 
of $\nu N$ DIS  and 
rely on the appearance of the future Neutrino Factory, 
which may provide data for the Bjunp sum rule as well  (for a  discussion of 
this possibility, see Refs. \cite{Mangano:2001mj,Alekhin:2002pj}).

{\bf Acknowkedgements}
It is a pleasure to thank D.J. Broadhurst for a  productive 
collaboration, which allowed me to broaden  my understanding 
of the theoretical structure of different DIS sum rules.
I am grateful to S.I. Alekhin, G. Altarelli, Yu.L. Dokshitzer and  
V.I. Zakharov for discussions. This work is supported by RFBR Grants 
N 03-02-17047, 03-02-17177 and is done within the scientific 
program of RFBR grant N 05-01-00992. The work was completed  
during  visit to CERN. I have a pleasure in thanking  the 
members of the CERN  Theory Group for encouraging hospitality.

\end{document}